\documentstyle[preprint,aps]{revtex}
\input epsf.tex
\def\DESepsf(#1 width #2){\epsfxsize=#2 \epsfbox{#1}}
\begin{document}
\preprint{\vbox{\hbox{OITS-560}}}
\draft
\title{Unique Signature of Electroweak Penguin\\
In Pure Hadronic $B$ Decays
\footnote{Work supported in part by the Department of Energy Grant No.
DE-FG06-85ER40224.}}
\author{N.G. Deshpande, Xiao-Gang He and Josip Trampetic\footnote{On leave of
absence from Department of Theoretical Physics, R. Boskovic Institute, Zagreb,
Croatia.}}
\address{Institute of Theoretical Science\\
University of Oregon\\
Eugene, OR 97403-5203, USA}

\date{October 1994}
\maketitle
\begin{abstract}
We study elelctroweak penguin contributions in pure hadronic $B$ decays in the
Standard and two-Higgs-doublet models.
We find that in $B_s\rightarrow \pi^0 (\eta\;, \phi)$ and  $B_s\rightarrow
\rho^0 (\eta\;, \phi)$ decays, the electroweak penguin contributions dominate
over
all other contribuitons. These processes provide unique signatures of
electroweak penguin in pure hadronic processes. These modes could be in
measureable range of future $B$ facilities.
\end{abstract}
$B$ decays are very interesting processes to study. They provide tests for the
Standard Model (SM) at both the tree and loop levels\cite{stone}. In the SM,
radiative and leptonic $B^0$ decays proceed through penguin loop inteactions.
Radiative
$b\rightarrow s \gamma$ decay has recently been observed at CLEO\cite{cleo}.
The measured branching ratio is consistent with the SM prediction\cite{bsg}.
The leptonic decays have
not yet been observed.
For hadronic $B$ decays, the effective Hamiltonian
contains the tree, the strong penguin and the electroweak penguin interactions.
In charmed $B$ decays, the tree level contributions are the dominant
ones\cite{stech,gatto}. In charmless $B$ decyas with $\Delta S = 0$, the tree
contributions are again the dominant ones. The strong and electroweak penguin
contributions are less than 20\%\cite{dh2}. In charmless $B$ decays with
$\Delta S = \pm 1$, the
tree contributions are proportional to $V_{ub}V^*_{us}$, wherease the strong
penguin contributions are proportional to $V_{tb}V^*_{ts}$. The KM enhancement
factor for the penguin decays is large $|V_{tb}V^*_{ts}|/|V_{ub}V_{us}^*|
\approx 50$. Although the Wilson coefficients (WC) for the penguins are much
smaller
than those of the tree WCs, the contributions from the strong penguin
dominate\cite{dt1}.
Processes mediated by strong penguin provide tests for the SM at the loop
level.
Naively one would think that the electroweak penguin contributions are
suppressed by a factor of $\alpha_{em}/\alpha_s$ compared with the strong
penguin
contributions, and therefore can be negelcted. However this is not true for
large top quark mass, and the electroweak
penguin contributions can be significant. It was found that in pure $\Delta S =
1$ penguin decays, the electroweak penguin contribuiton can reduce the
branching ratios by 20\% to 30\%\cite{dh2}. The large electroweak penguin
contributions have
important implications for many decay processes\cite{dh1}. In some processes
the electroweak penguin effects could, in fact, dominate.
In this paper we point out that there are hadronic processes in which the
electroweak penguin contributions, indeed, dominate. These processes therefore
provide unique signatures for electroweak penguin effects in pure hadronic
decays.

In $\Delta S = 1$ $B$ decays, the strong penguins are pure $\Delta I
= 0$ operators, whereas the tree and electroweak penguin operators contain both
$\Delta I = 0$
and $1$ components. In order to find possible processes where electroweak
penguins dominate, it is clear that one should look for processes which are
$\Delta I = 1$ such that the strong penguins will not contribute at all. For
two body decays, we have found four such processes, namely, $\bar B_s^0
\rightarrow \eta\pi^0\;, \eta \rho^0\;,
\phi \pi^0\;, \phi \rho^0$. One of these modes, $\bar B_s^0 \rightarrow \phi
\pi^0$, has been discussed previously in Ref.\cite{add} to leading order. The
tree level contribuitons are highly suppressed because they are proportional to
$|V_{ub}V^*_{us}|$. Furhtermore, they are color suppressed. Using factorization
approximation, we find that in the SM  for $N_c = 3$,
the electroweak penguin contributions enhance the branching ratios by about
a factor of 100 compared to the tree contributions. For $N_c = 2$, the
enhancement factor can still be about 10. The electroweak contributions
dominate and the branching ratios are much larger than previous
estimates\cite{gatto}. We have also studied these processes in the two Higgs
doublet models. We find that the enhancement factors can be bigger for a range
of $m_H$ and $cot\beta$.

The QCD corrected Hamiltonian relevant to us
can be written as follows\cite{buras}:
\begin{eqnarray}
H_{eff} = {G_F\over \sqrt{2}}[V_{ub}V^*_{us}(c_1O_1 + c_2 O_2) -
V_{tb}V^*_{ts}\sum_{i=3}^{10} c_iO_i] +H.C.\;,
\end{eqnarray}
where the Wilson coefficients (WCs) $c_i$ are defined at the scale of $\mu
\approx
m_b$; and $O_i$ are defined as
\begin{eqnarray}
O_1 &=& \bar s_\alpha \gamma_\mu(1-\gamma_5)u_\beta\bar
u_\beta\gamma^\mu(1-\gamma_5)b_\alpha\;,\;O_2 =\bar s
\gamma_\mu(1-\gamma_5)u\bar
u\gamma^\mu(1-\gamma_5)b\;,\nonumber\\
O_3 &=&\bar s \gamma_\mu(1-\gamma_5)b
\bar q' \gamma_\mu(1-\gamma_5) q'\;,\;\;\;\;\;\;\;O_4 = \bar s_\alpha
\gamma_\mu(1-\gamma_5)b_\beta
\bar q'_\beta \gamma_\mu(1-\gamma_5) q'_\alpha\;,\\
O_5 &=&\bar s \gamma_\mu(1-\gamma_5)b  \bar q'
\gamma^\mu(1+\gamma_5)q'\;,\;\;\;\;\;\;\;O_6 = \bar s_\alpha
\gamma_\mu(1-\gamma_5)b_\beta
\bar q'_\beta \gamma_\mu(1+\gamma_5) q'_\alpha\;,\nonumber\\
O_7 &=& {3\over 2}\bar s \gamma_\mu(1-\gamma_5)b  e_{q'}\bar q'
\gamma^\mu(1+\gamma_5)q'\;,\;O_8 = {3\over 2}\bar s_\alpha
\gamma_\mu(1-\gamma_5)b_\beta
e_{q'}\bar q'_\beta \gamma_\mu(1+\gamma_5) q'_\alpha\;,\nonumber\\
O_9 &=& {3\over 2}\bar s \gamma_\mu(1-\gamma_5)b  e_{q'}\bar q'
\gamma^\mu(1-\gamma_5)q'\;,\;O_{10} = {3\over 2}\bar s_\alpha
\gamma_\mu(1-\gamma_5)b_\beta
e_{q'}\bar q'_\beta \gamma_\mu(1-\gamma_5) q'_\alpha\;.\nonumber
\end{eqnarray}
Here $q'$ is summed over u, d, and s.

We work with renormalization scheme independent WCs $c_i$ to next-to-leading
order in QCD correstions as discussed in
Ref.\cite{dh1,buras}. We use the top quark mass as indicated by the CDF
measurement of $m_t = 174$ GeV\cite{cdf} and the world average value for the
strong coupling constant $\alpha_s(m_Z^2) = 0.117$\cite{particle}, we have
\begin{eqnarray}
&c_1 = -0.307\;, \;\;\;c_2 = 1.147\;,\;\;\; c_3 = 1.71\times 10^{-2}\;,\;\;\;
c_4 = -3.68\times^{-2}\;,\nonumber\\
&c_5 = 1.03\times 10^{-2}\;,\;\;\;
c_6 = -4.50\times 10^{-2}\;, \;\;\;c_7 = -1.24\times 10^{-5}\;, \nonumber\\
&c_8 = 3.77\times10^{-4}\;,\;\;\;
c_9 = -1.03\times 10^{-2}\;,\;\;\; c_{10} = 2.06\times 10^{-3}\;.
\end{eqnarray}

We need to treat the matrix elements to one-loop level for consistency.
These one-loop matrix elements can be rewritten in terms of the tree-level
matrix elements $<O_j>^{t}$ of the effective operators, and one finds
\cite{dh1,fleischer} $<c_iQ_i>$ to be equal to
\begin{eqnarray} c_i [\delta_{ij}+{\alpha_s\over 4\pi}m^s_{ij}
+{\alpha_{em}\over 4\pi}m^e_{ij}] <O_j>^{t}
\equiv c_i^{eff}<O_i>^{t}.
\end{eqnarray}
We have worked out the full matrices $m^{s,e}$. To the order we are
considering only $ c_{3-10}$ will change.
They are given by,
\begin{eqnarray}
&c^{eff}_3 = c_3 - P_s/3\;,\;\; c^{eff}_4 = c_4 +P_s\;,\;\;\;
c^{eff}_5 = c_5 - P_s/3\;,\nonumber\\
&c^{eff}_6 = c_6 + P_s\;,
c^{eff}_7 = c_7 +P_e\;,\;\;\;c_8^{eff} =
c_8\;,\nonumber\\
&c_9^{eff} =  c_9 +P_e\;,\;\;\; c_{10}^{eff} =  c_{10}\;.
\end{eqnarray}
The leading contributions to $P_{s,e}$ are given by:
 $P_s = (\alpha_s/8\pi) c_2 (10/9 +G(m_c,\mu,q^2))$ and
$P_e = (\alpha_{em}/9\pi)(3 c_1+ c_2) (10/9 + G(m_c,\mu,q^2))$. Here
$m_c$ is the charm quark mass which we take to be 1.35 GeV. The function
$G(m,\mu,q^2)$ is give by
\begin{eqnarray}
G(m,\mu,q^2) = 4\int^1_0 x(1-x) \mbox{d}x \mbox{ln}{m^2-x(1-x)q^2\over
\mu^2}\;.
\end{eqnarray}
In the numerical calculation, we will use $q^2 = m_b^2/2$ which represents the
average value.

Using factorization approximation we find
\begin{eqnarray}
<\eta \pi^0|H_{eff}|\bar B^0_s> = {G_F\over \sqrt{2}}C_\pi<\pi^0|\bar
u\gamma_\mu \gamma_5u|0><\eta|\bar s \gamma^\mu(1-\gamma_5)b|\bar
B^0_s>\nonumber\;,\\
<\eta \rho^0|H_{eff}|\bar B^0_s> = {G_F\over \sqrt{2}}C_\rho<\rho^0|\bar
u\gamma_\mu  u|0><\eta|\bar s \gamma^\mu(1-\gamma_5)b|\bar B^0_s>\nonumber\;,\\
<\phi \pi^0|H_{eff}|\bar B^0_s> = {G_F\over \sqrt{2}}C_\pi<\pi^0|\bar
u\gamma_\mu \gamma_5 u|0><\phi|\bar s \gamma^\mu(1-\gamma_5)b|\bar
B^0_s>\nonumber\;,\\
<\phi \rho^0|H_{eff}|\bar B^0_s> = {G_F\over \sqrt{2}}C_\rho<\rho^0|\bar
u\gamma_\mu  u|0><\phi|\bar s \gamma^\mu(1-\gamma_5)b|\bar B^0_s>\nonumber\;,
\end{eqnarray}
where $C_{\pi} = V_{ub}V_{us}^*(c^{eff}_1
+\xi c^{eff}_2) - V_{tb}V_{ts}^* {3\over 2}(-c^{eff}_7 - \xi c^{eff}_8
+c_9^{eff}+\xi c_{10}^{eff})$. Here $\xi = 1/N_c$.  $C_\rho$ is obtained by
replacing $-(c^{eff}_{7}+\xi c^{eff}_8)$ by
$+(c^{eff}_{7}+\xi c^{eff}_8)$.

In order to obtain the branching ratios we need to evaluate the hadronic matrix
elements. We use the following Lorentz decomposition for the relevant current
matrix elements,
\begin{eqnarray}
&<&\pi^0(q)|\bar u\gamma_\mu(1-\gamma_5)u|0> = -if_{\pi^0} q_\mu\;,\;\;
<\rho^0(q)|\bar u\gamma_\mu u|0> = i f_{\rho^0} \epsilon(q)_\mu\;,\\
&<&\eta(k)|\bar s \gamma^\mu(1-\gamma_5)b|\bar B^0_s(p)> = {2\over\sqrt{6}}
[(p+k)^\mu F_1(q^2)
- {m_\eta^2-m_B^2\over q^2}q^\mu (F_0(q^2)-F_1(q^2))]\;,\nonumber\\
&<&\phi(k)|\bar s \gamma^\mu(1-\gamma_5)b|\bar B^0_s(p)> =
i \epsilon^{\mu\nu\lambda\sigma}\epsilon_\nu(k) (k+p)_\lambda (p-k)_\sigma
V(q^2)\nonumber\\
&+&(m_{B_s}^2-m_\rho^2)\epsilon^\mu(k)A_1(q^2)
-(\epsilon(k)\cdot q) (k+p)^\mu A_2(q^2)
+(\epsilon(k)\cdot q) (m_{B_s}+m_\rho){q^\mu\over q^2}\nonumber\\
&\times& [2m_\rho A_0(q^2)
+(m_{B_s}-m_\rho)(A_2(q^2)-A_1(q^2))]\;,\nonumber
\end{eqnarray}
where $q=p-k$. The $f_{\pi^0}
= 93\; \mbox{MeV}$ is the pion decay constant, and the $f_{\rho^0} = 156\;
\mbox{MeV}^2$ is the $\rho$-meson decay constant. The form factors
$F_{0,1}(q^2)$ are the same as those defined in Ref.\cite{stech}, while
$V(q^2)\;, A_{0,1,2}(q^2)$ are related to the from factors $V(0)\;,
A_{0,1,2}(0)$ defined in Ref.\cite{stech} by
\begin{eqnarray}
V(q^2) = {1\over (m_{B_s}+m_\phi)}{V(0)\over 1-q^2/m_{1^-}^2}\;&,&
A_0(q^2) = {1\over (m_{B_s}+m_\phi)}{A_0(0)\over 1-q^2/m_{0^-}^2}\;,\nonumber\\
A_1(q^2)= {1\over (m_{B_s}-m_\phi)}{A_1(0)\over 1-q^2/m_{1^+}^2}\;&,&
A_2(q^2) = {1\over (m_{B_s}+m_\phi)}{A_2(0)\over 1-q^2/m_{1^+}^2}\;,
\end{eqnarray}
The masses of the excited meson states, corresponding to the $\bar s b$
current, are: $m_{0^-} = 5.38$ GeV,  $m_{1^-} = 5.43$ GeV,  $m_{0^-} = 5.89$
GeV,  $m_{1^+} = 5.82$ GeV. In our numerical calculations, we will use the form
factors at $q^2=0$ estimated in Ref.\cite{gatto1}.
Normalizing the branching ratios to the
tree contribution $BR_{tree}$, we write the total branching ratios $BR$ as
\begin{eqnarray}
BR(\bar B_s^0\rightarrow \pi^0 (\eta, \phi)) &=& BR_{tree}(\bar
B_s^0\rightarrow \pi^0 (\eta, \phi))|1 + R_\pi|^2\;,\nonumber\\
BR(\bar B_s^0\rightarrow \rho^0 (\eta, \phi)) &=& BR_{tree}(\bar
B_s^0\rightarrow \rho^0 (\eta, \phi))|1 + R_\rho|^2\;,
\end{eqnarray}
where $R_{\pi(\rho)} = - [V_{tb}V_{ts}^* {3\over 2}(-(+)(c^{eff}_7 + \xi
c^{eff}_8)
+c_9^{eff}+\xi c_{10}^{eff}]/(V_{ub}V_{us}^*(c^{eff}_1
+\xi c^{eff}_2))$, and $BR_{tree}$'s can be obtained from the tree decay widths
\begin{eqnarray}
&\Gamma_{tree}&(\bar B_s^0\rightarrow \pi^0 \eta) = {G_F^2F_0^2(m_\pi^2)
f^2_{\pi^0}\over 48\pi}|C_{tree}|^2\lambda_{\eta\pi}m_{B_s}^3(1-{m_\eta^2\over
m_{B_s}^2})^2\;,\nonumber\\
&\Gamma_{tree}&(\bar B_s^0\rightarrow \pi^0 \phi)=
{G_F^2(m_{B_s}+m_\phi)^2A_0^2(m_\pi^2)
f^2_{\pi^0}\over
32\pi}|C_{tree}|^2\lambda_{\phi\pi}^{3/2}m_{B_s}^3\;,\nonumber\\
&\Gamma_{tree}&(\bar B_s^0\rightarrow \rho^0 \eta) = {G_F^2F_1^2(m_\rho^2)
f^2_{\rho^0}\over 48\pi
m_\rho^2}|C_{tree}|^2\lambda_{\eta\rho}^{3/2}m_{B_s}^3\;,\nonumber\\
&\Gamma_{tree}&(\bar B_s^0\rightarrow \rho^0 \phi)= {G_F^2
f^2_{\rho^0}\over 32\pi}|C_{tree}|^2\lambda_{\phi\rho}^{3/2}m_{B_s}^3
 [2V^2(m_\rho^2)
+ {3\over \lambda_{\phi\rho}}(1- {m_\phi^2\over m_{B_s}^2})^2 A_1^2(m_\rho^2)
\nonumber\\
&-& A_2^2(m_\rho^2)
+{m_{B_s}^4\over 4m_\rho^2 m_\phi^2} ((1-{m_\phi^2\over
m_{B_s}^2})A_1(m_\rho^2)
-(1-{m_\phi^2+m_\rho^2\over m_{B_s}^2})A_2(m^2_\rho))^2]\;,
\end{eqnarray}
where $\lambda_{ab} = \sqrt{1- 2(m_a^2+m_b^2)/m_{B_s}^2 +
(m_a^2-m_b^2)^2/m_{B_s}^4}$, and $C_{tree} = V_{ub}V_{us}^*(c_1^{eff}+\xi
c^{eff}_2)$.

Using
the Wolfenstein parametrization for the KM matrix, and
$|V_{ts}/V_{ub}|\approx 1/0.08$, $V_{us} = 0.22$, we have
\begin{eqnarray}
R_{\pi(\rho)} = 85 e^{i\gamma}{-(+)(c_7^{eff}+\xi c_8^{eff}) +c_9^{eff}+\xi
c_{10}^{eff}
\over c_1 +\xi c_2}\;.
\end{eqnarray}
Using the WCs given before, for $N_c = 3$ we have
\begin{eqnarray}
BR(\bar B_s^0\rightarrow \pi^0 (\eta, \phi)) &=& BR_{tree}(\bar
B_s^0\rightarrow \pi^0 (\eta, \phi))|1 - 11.1e^{i\gamma}|^2\;,\nonumber\\
BR(\bar B_s^0\rightarrow \rho^0 (\eta, \phi)) &=& BR_{tree}(\bar
B_s^0\rightarrow \pi^0 (\eta, \phi))|1 - 10.9e^{i\gamma}|^2\;,
\end{eqnarray}
 And for $N_c = 2$ we have
\begin{eqnarray}
BR(\bar B_s^0\rightarrow \pi^0 (\eta, \phi)) &=& BR_{tree}(\bar
B_s^0\rightarrow \pi^0 (\eta, \phi))|1 - 3.03e^{i\gamma}|^2\;,\nonumber\\
BR(\bar B_s^0\rightarrow \rho^0 (\eta, \phi)) &=& BR_{tree}(\bar
B_s^0\rightarrow \pi^0 (\eta, \phi))|1 - 2.95e^{i\gamma}|^2\;,
\end{eqnarray}
In the above equations, we have neglected a small imaginary part for $\bar
B^0_s
\rightarrow \rho^0(\eta\;,\phi)$.
We clearly see the dominance of the electroweak penguin contributions.
For $N_c = 3$, the enhancement factor in the branching ratios is about
100. For $N_c = 2$, the enhancement factor is between 4 to 16 depending on the
weak phase $\gamma$. The tree level contribution is proportional to
$a_2 = c_1 +\xi c_2$. The determination of $a_2$ from experimental data
indicates
that $N_c$ is close to 2\cite{gatto1}. So the enhancement factor is likely to
be between 4 to
16. Because the dominant contributions are from the electroweak penguin
contributions, the total branching ratios are not sensitive to $N_c$.

For $N_c = 2$, we obtain the following branching ratios:
\begin{eqnarray}
BR(\bar B_s \rightarrow \pi^0\eta) = 2.1 \times
10^{-8}|1-3.03e^{i\gamma}|^2\;,\;\;
BR(\bar B_s \rightarrow \rho^0\eta) = 5.8 \times
10^{-8}|1-2.95e^{i\gamma}|^2\;,\nonumber\\
BR(\bar B_s \rightarrow \pi^0\phi) = 3.1 \times
10^{-9}|1-3.03e^{i\gamma}|^2\;,\;\;
BR(\bar B_s \rightarrow \rho^0\phi) = 1.4 \times
10^{-8}|1-2.95e^{i\gamma}|^2\;.
\end{eqnarray}
Without the electroweak penguin contributions they are all small ($< 10^{-7}$).
With electroweak penguin contributions, they can be
enhanced by about one order of magnitude. The branching ratio for $\bar B_s
\rightarrow \rho^0\eta$ can be as large as $10^{-6}$. These branching ratios,
although small, may be accessible at a hadron collider that have large b quark
production.

We now discuss possible new contributions in the two Higgs doublet models.
In the two-Higgs-doublet model, there are new contributions due to charged
Higgs boson. The charged Higgs-quark couplings are given
by
\cite{higgs}
\begin{eqnarray}
L_H = {g\over 2\sqrt{2}m_W} \bar u_i V_{ij}
[cot\beta m_{u_i}(1-\gamma_5)- a m_{d_j}(1+\gamma_5)]d_j H^+
+H.C.\;,
\end{eqnarray}
where $cot\beta = v_1/v_2$; $v_1$ and $v_2$ are the vacuum expectation values
of the Higgs doublets $H_1$ and $H_2$,  which generate masses for down
and up quarks, respectively. The parameter $a$ depends on the
models\cite{higgs}. There are two ways of providing masses to the quarks from
Higgs doublets. One
of these is to generate up quark masses by the vaccum expectation value $v_2$
of $H_2$, and the down quark masses by $v_1$ of $H_1$ (Model I). The minimal
supersymmetric SM is such a model. The other is to generate all quark masses
from one Higgs doublet, for example $H_2$. For Model I, $a = tan\beta$, and
Model II, $a = cot\beta$. We use the recent experimental data on $b\rightarrow
s \gamma$ from CLEO to constrain the allowed regions for the charged Higgs mass
$m_H$ vs. $cot\beta$\cite{cleo}. Although the additional contributions due to
charged Higgs boson come in the same form for the two models, the allowed
regions are different. We present our results for $R_{\pi(\rho)}$ in Figure 1.

We see from the figure that the decays $\bar B_s \rightarrow \pi^0 (\eta,
\phi)$
are more sensitive to charged Higgs effects than  $\bar B_s \rightarrow \rho^0
(\eta, \phi)$.
Experimental constraint from $b\rightarrow s \gamma$ is very strigent for
Model I. The Higgs mass is constrained to be larger than 260 GeV. For $m_H
=500$
GeV, $cot\beta$ is constrained to be less than 1, the charged Higgs effect is
less than 10\% in R for  $\bar B_s \rightarrow \pi^0 (\eta, \phi))$. For
 $\bar B_s \rightarrow \rho^0 (\eta, \phi))$ the charged Higgs effect is even
smaller. For $m_H = 1$ TeV, $cot\beta$ can be about 3. The charged Higgs effect
can enhance R by about 30\% for  $\bar B_s \rightarrow \pi^0 (\eta, \phi))$.
For
 $\bar B_s \rightarrow \rho^0 (\eta, \phi))$ the effect is less than 10\%.
For Model II, the constraints on the parameters are less stringent. For
the charged Higgs in the range of 100 GeV to 1 TeV, R can be as large as 7 for
 $\bar B_s \rightarrow \pi^0 (\eta, \phi))$ decays. The branching ratios can be
enhanced by about 40 to 60 times compared with the tree level branching ratios.
The R factor for
$\bar B_s \rightarrow \rho^0 (\eta, \phi))$ decays is smaller, but it can be as
large as 5. The branching ratios can be enhanced by about 15 to 36 times.

In summary, we have found four processes where electroweak penguin effects
dominate. These processes were highly suppressed without the penguin
contributions, but could now be in the range of a future $B$ facility, probably
a hadron facility with large $b$ production. Effects of significant
enhancements from non standard physics can also be looked for, and useful
limits might be set
on masses of new particles.

\begin{figure}[htb]
\centerline{ \DESepsf(fig1.epsf width 10 cm) }
\smallskip
\caption{$R_{\pi(\rho)}$ as a function of $cot\beta$ with
$m_t = 174$ GeV, $N_c = 2$,
and $\alpha_s(m_Z) = 0.117$ for different charged Higgs masses. The solid lines
are for $\bar B_s \rightarrow \pi^0(\eta\;, \phi)$, and the dashed lines are
for $\bar B_s \rightarrow \rho^0(\eta\;, \phi)$. The curves 1, 2 ,3 are for
$m_H = (100\;, 500\;, 1000)$ GeV, respectively.}
\label{fig1}
\end{figure}

\end{document}